# Colossal optical anisotropy in wide-bandgap semiconductor $CuAlO_2$

Baekjune Kang,[1†] Junhee Shin,[1†] Myeongjun Kang,[2] Uksam Choi,[1] Uihyeon Seo,[1] Kunook Chung,[1] Jong Mok Ok,[2*] Hosub Jin,[1*] and Changhee Sohn[1*]

[1]*Department of Physics, Ulsan National Institute of Science and Technology, Ulsan, 44919, Republic of Korea*

[2]*Department of Physics, Pusan National University, Pusan, 46241, Republic of Korea*

**Abstract**

Colossal optical anisotropy in the entire visible spectrum is crucial for advanced photonic applications, enabling precise light manipulation without optical loss across a broad spectral range. Here, we demonstrate that $CuAlO_2$ exhibits colossal optical anisotropy and transparency across the visible spectrum, enabled by its unique three-dimensional O-Cu-O dumbbell structure and two-dimensionally confined excitons. Using mm-sized single crystals, we independently measured *ab*-plane and *c*-axis optical properties, revealing maximum birefringence ($|\Delta n_{max}| = 3.67$) and linear dichroism ($|\Delta k_{max}| = 5.21$), the highest reported to date. $CuAlO_2$ retains birefringence over 0.5 throughout the entire visible range and possesses a wide direct bandgap of 3.71 eV, surpassing the birefringence of commercial anisotropic crystals transparent in the visible spectrum. From the two-dimensional screened hydrogen model and first-principles calculations, we demonstrate that the colossal anisotropy arises from a unique excitonic Cu *d-p* transition confined to the atomic-thick layer. This colossal optical anisotropy and transparency across the entire visible spectrum makes $CuAlO_2$ a promising candidate for future photonic technologies.

**Introduction**

Materials with both large optical anisotropy and wide bandgaps are extremely rare in nature but essential for advanced photonic technologies. Optical anisotropy allows light to propagate differently depending on its polarization and the material's orientation, making it critical for applications such as optical communications, phase-matching elements, and polarizers[1-4]. However, these devices typically operate with light energies below the material's bandgap to minimize optical loss, presenting a significant challenge in material selection. While van der Waals materials have garnered considerable attention for their exceptional optical anisotropy[5], their interband transitions below 2 eV limit their applicability in the visible range[6]. On the other hand, anisotropic materials like $LiNbO_3$ that maintain transparency up to the UV range typically possess only modest anisotropy in this spectral region[7]. This trade-off has left researchers grappling with a persistent dilemma: strong optical anisotropy is often accompanied by narrow bandgaps, while wider bandgaps tend to weaken anisotropy. Although metamaterials have made promising progress in addressing these limitations, issues like scattering losses at interfaces and manufacturing complexities remain hurdles to widespread use[8]. In light of these limitations, there is an urgent need to identify natural materials that combine giant optical anisotropy with wide bandgaps[9-11].

Delafossite $CuAlO_2$ offers a promising platform for realizing significant anisotropic optical properties in the UV-visible range. $CuAlO_2$ consists of alternating $Cu^+$ layers and $AlO_2^-$ layers along the *c*-axis (Fig. 1a). The $Cu^+$ layer creates an O-Cu-O dumbbell structure along the *c*-axis and a triangular lattice composed of Cu atoms in the *ab*-plane. This structural anisotropy suggests that if the optical response is dominated by the $Cu^+$ ions, strong anisotropic optical properties can emerge. Indeed, previous theoretical studies have suggested that the optical response of $CuAlO_2$ is governed by quasi-two-dimensional excitons in $Cu^+$ layers[12,13], implying giant optical anisotropy. In addition, as a *p*-type wide-bandgap semicondcutor[14], it exhibits a direct bandgap of ~3.5 eV and an indirect bandgap of ~3.0 eV[15], providing excellent transparency in the up to UV range. However, investigation of its anisotropic optical behavior and exciton in the UV-visible range has been limited by the difficulty of growing sufficiently large single crystals of $CuAlO_2$[16,17].

Here, we observe colossal optical anisotropy and transparency of $CuAlO_2$ across the entire visible spectrum. Using a new synthesis method[18], we produce mm-sized $CuAlO_2$ single crystals (Supplementary Note 1) with smooth and shiny side faces, allowing independent measurement of both *ab*-plane (top surface) and *c*-axis (side surface) optical properties without any model fitting (Supplementary Note 2). Unlike van der Waals materials, $CuAlO_2$'s three-dimensional crystal structure formed by the O-Cu-O dumbbell configuration allows for sufficient thickness and ensures practical utilization of out-of-plane anisotropy (Fig. 1b). Our measurements reveal maximum values of birefringence ($|\Delta n_{max}|$ = 3.67) and linear dichroism ($|\Delta k_{max}|$ = 5.21), both of which represent the highest values reported to date for any material in UV-visible range. $CuAlO_2$ also shows a wide direct optical bandgap of 3.71 eV (Supplementary Note 3) and retains $|\Delta n(\omega)|$ exceeding 0.5 across the entire visible spectrum, surpassing the $|\Delta n(\omega)|$ of commercially available anisotropic crystals transparent in the visible range. Through a thorough investigation of optical spectra and first-principles calculation, we revealed the origin of colossal anisotropy in $CuAlO_2$ is a two-dimensionally confined exciton despite a three-dimensional crystal structure. As shown in Fig 1c, these excitonic transitions are only allowed for light polarized in *ab*-plane, resulting in exceptional optical anisotropy in $CuAlO_2$.

**Results**

Figures 2a and 2b are the real ($\varepsilon_1(\omega)$) and imaginary ($\varepsilon_2(\omega)$) parts of the dielectric function obtained from the ellipsometry parameters (method). The $\varepsilon_1(\omega)$ and $\varepsilon_2(\omega)$ with light polarization parallel to *ab*-plane (red) and *c*-axis (blue) of $CuAlO_2$ are remarkably different from each other, indicating significant optical anisotropy. In particular, an extremely sharp and intense optical transition around 3.78 eV exclusively occurs with light polarization parallel to the *ab*-plane while nearly zero spectral weight is observed with light polarization parallel to the *c*-axis below 5 eV. Such discrepancy in this optical transition is the major source of optical anisotropy in UV-visible wavelength. The unusually large and sharp peak in $\varepsilon_{ab2}(\omega)$ suggests that the optical response is dominated by excitons. This is because the strong electron-hole interaction significantly enhances oscillator strength and prolongs excited-state lifetimes[12]. Furthermore, the narrow linewidth of approximately 200 meV at room temperature suggests that the exciton exhibits characteristics typical of low-dimensional systems, an unusual feature for three-dimensional, mm-thickness bulk $CuAlO_2$[19].

The significant optical anisotropy and wide bandgap of $CuAlO_2$ make it suitable for photonic technologies that exploit colossal birefringence and low optical loss. Figure 2c-2f displays the refractive index ($n(\omega)$), optical extinction coefficient ($k(\omega)$), $|\Delta n(\omega)|$, $|\Delta k(\omega)|$, and absorption ($\alpha(\omega)$) of $CuAlO_2$ at room temperature. All results support the colossal optical anisotropy of $CuAlO_2$. The most intriguing property is its $|\Delta n(\omega)|$. The $|\Delta n(\omega)|$ exceeds 0.5 across nearly the entire range, except immediately after the optical transition (3.86 - 3.9 eV),

and even surpasses 1 between 2.86 eV and 6.2 eV. Notably, the maximum values of birefringence ($|\Delta n_{max}| = 3.67$) and refractive index ($n_{ab\,max} = 5.66$) appear near the peak. Additionally, $|\Delta k(\omega)|$ is observed to be very strong, with a maximum $|\Delta k_{max}|$ value of approximately 5.21, further emphasizing the significant optical anisotropy of $CuAlO_2$. These maximum $|\Delta n_{max}|$ and $|\Delta k_{max}|$ values are, to the best of the authors' knowledge, the highest reported in the UV-visible range. Unlike other highly anisotropic materials such as van der Walls materials, which typically exhibit absorption in the visible range[6], $CuAlO_2$ remains inherently transparent down to approximately 370 nm and possesses a direct bandgap of about 3.71 eV[15] (Fig. 2f and Supplementary Note 3).

Figure 3 compares the $|\Delta n(\omega)|$ of $CuAlO_2$ in its transparent range with other well-known anisotropic materials, such as rutile and $LiNbO_3$[5,7,20-25]. In this comparison, we selected materials that exhibit transparency below 1.5 eV (~ 800 nm) to ensure compatibility with a visible range of optical technologies. Here, we define the transparent region as the Urbach absorption tail below the direct bandgap[26], where defect-induced absorption occurs. The results highlight that $CuAlO_2$ not only exceeds the $|\Delta n(\omega)|$ of these conventional materials but also maintains its high value across a broader spectral range, spanning the entire visible range and extending into the UV range. Notably, $CuAlO_2$ achieves a maximum $|\Delta n(\omega)|$ value of 1.66 at 370 nm within its transparent region, which is the highest value for other UV range transparent anisotropic materials[27]. With its ability to sustain high $|\Delta n(\omega)|$ in a UV-visible range, $CuAlO_2$ emerges as a highly promising candidate for advanced photonic devices requiring efficient light manipulation across a wide spectrum, including the UV region.

**Discussion**

A natural question arising from the observed colossal optical anisotropy in $CuAlO_2$ is why such a sharp and strong exciton occurs exclusively for *ab*-plane polarized light. It is mainly attributed to the highly confined exciton in $Cu^+$ layers, resulting in quasi-two-dimensional exciton despite the material's three-dimensional crystal structures. The first signature of the two-dimensional nature of the exciton is 1/4 ratio of the direct electronic bandgap ($E_g$) to the binding energy ($E_b$). Previous theoretical and experimental studies have demonstrated that the 1/4 scaling relation between $E_b$ and $E_g$ is indicative of strongly bound excitons in two-dimensional materials[28-38] (Fig. 4a). To investigate this relation, we fitted the real part of the *ab*-plane dielectric function ($\varepsilon_{ab1}(\omega)$) using a model that accounts for the intense optical gap (Penn gap)[17,39]. Our analysis revealed an $E_g$ of 5.12 eV and an $E_b$ of 1.34 eV (Supplementary Note 4), consistent with the 1/4 scaling relation. The obtained $E_g$ is in accord with the previous theoretical research of the electronic bandgap from self-consistent GW (sc-GW) calculation (5.1 eV), which is the most accurate *ab initio* method to predict electronic bandgap of wide bandgap semiconductor[40]. Furthermore, the remarkably large $E_b$ of 1.34 eV aligns with recent theoretical research predicting $E_b$ as 1.2 eV[13].

The second notable feature is the appearance of a higher-order exciton spectrum between the 1*s* exciton at 3.78 eV and the $E_g$ at 5.12 eV, which is well described by a two-dimensional model. In the $\varepsilon_{ab2}(\omega)$, a distinct and broad peak appears within this range, which can be attributed to the contributions from the 2*s* and higher-order exciton states. To analyze these states in detail, we simultaneously performed a Fano-Lorentz fitting of the $\varepsilon_{ab1}(\omega)$ and $\varepsilon_{ab2}(\omega)$. The Fano-Lorentz model was chosen over the conventional Lorentz model to capture the distinct Fano resonance behavior observed at low temperatures (Supplementary Note 5). Figure 4b presents the Fano-Lorentz fitting results of the dielectric function for CuAlO$_2$ at room temperature, showing that the peak positions of the 1s exciton (associated with the Fano term) and the 2*s*/higher-order excitons (Lorentz term) are located at 3.78 eV and 4.47 eV, respectively.

We confirmed that the two-dimensional exciton model effectively explains our observed 2*s* and higher-order exciton states at 4.47 eV through a comparison between the two-dimensional and three-dimensional models. Understanding higher-order excitons requires consideration of dimensionality and screening effects, as the Coulomb potential primarily determines the exciton wavefunction[41-43]. To identify the most suitable model for CuAlO$_2$, we compared the two-dimensional screened hydrogen model (2D model) with the three-dimensional isotropic hydrogen model (3D model). The 2D model was derived using the effective medium approximation[44-46] (Supplementary Note 6). Figures 4c (2D model) and 4d (3D model) display the simulated $\varepsilon_{ab2}(\omega)$ with calculated peak positions and oscillator strengths for the *n*-th exciton series, assuming identical linewidths. In the 2D model, the 2*s* exciton appears at 4.40 eV, effectively reproducing the broad continuum observed at 4.47 eV in the experimental results. Conversely, the 2*s* exciton in the 3D model emerges at 4.84 eV with a complete absence of spectral weight at 4.47 eV, indicating that the 3D model fails to capture this continuum (Supplementary note 7). These results conclusively demonstrate that the $\varepsilon_{ab2}(\omega)$ of CuAlO$_2$ is best described by the 2D model, reinforcing their quasi-two-dimensional nature of exciton in CuAlO$_2$.

The two-dimensionally confined exciton in CuAlO$_2$ can be effectively explained by its unique lattice structure and symmetry of band structure. Previous theoretical studies have identified quasi-two-dimensional optical transitions in CuAlO$_2$, suggesting that this behavior likely arises from its layered structure[12,13]. However, the specific mechanisms behind quasi-two-dimensional optical transitions have not been fully explored. To investigate the origin of two-dimensional exciton in more detail, we performed *ab initio* calculation using the generalized gradient approximation (GGA) method. GGA effectively captures the orbital characters arising from the material's symmetry despite its inaccurate bandgap, which lies outside the scope of our focus. Indeed, more complicated sc-GW calculations have shown that wavefunction corrections at the *L*-point where optical transition occurs[13,47,48] are minor except bandgap[49], justifying the use of GGA for understanding the optical properties and selection rules.

The distinct anisotropic optical transition processes in CuAlO$_2$ arise fundamentally from its orbital character and lattice structure. Figure 5a displays the band structure and partial density of states (PDOS) of CuAlO$_2$, showing qualitative consistency with previous research[12,50]. It highlights the major contributions near the Fermi level from Cu and O orbitals, with negligible contribution from Al orbitals. To understand the role of specific orbitals, we present the orbital projected band structure and PDOS with Cu *d* orbitals (Fig. 5b), Cu *p* orbitals (Fig. 5c), and O *p* orbitals (Fig. 5d). The highest valence band and lowest conduction band are dominated by Cu $d_{z^2}$ orbital character, with significant contributions from Cu $p_{x/y}$ in the conduction band. As indicated by a black arrow and red box in Fig. 5a, the lowest and primary optical transition occurs along the *L-F* line. Here, the $d_{z^2}$ character vanishes entirely in the conduction band (Fig. 5b) while Cu $p_{x/y}$ orbitals dominate. The complete separation of Cu 3*d* and 4*p* orbitals is due to the forbidden *d-p* intratom orbital mixing within the linear coordinate of the O-Cu-O dumbbell structure[51]. Furthermore, this dumbbell structure locates the Cu $p_z$ orbital with higher energy compared to the Cu $p_{x/y}$ orbital. As a result, the lowest and primary optical excitation becomes intratom-like from Cu $d_{z^2}$ orbital to Cu $p_{x/y}$ orbitals.

These results are confirmed by the Bloch functions, obtained from the highest valence band and lowest conduction band at the *L*-point (Fig. 5e). The Bloch function of the valence band shows a dominant Cu $d_{z^2}$ with minor O $p_z$ extending toward the Cu atom. In contrast, the Bloch function of the conduction band displays the major Cu $p_{x/y}$ with minor O $p_{x/y}$ orbitals stretching toward the Al atom. Based on selection rules, transitions induced by *ab*-plane polarized light are allowed the transition from Cu $d_{z^2}$ orbital to Cu $p_{x/y}$ orbital, while transitions occurring by *c*-axis polarized light are forbidden[52]. Consequently, the O-Cu-O dumbbell structure and orbital character are the origins of the experimentally observed colossal optical anisotropy.

Our experimental and theoretical studies demonstrate the colossal optical anisotropy and two-dimensional exciton behavior of $CuAlO_2$. These unique optical properties arise from $CuAlO_2$'s anisotropic lattice structure and specific symmetries in the Brillouin zone. Importantly, our findings extend beyond $CuAlO_2$ and apply to other materials with dumbbell structures. Since the O-Cu-O dumbbell structure is the main origin of colossal optical anisotropy, the other delafossite or new layered materials with dumbbell structure can provide the possibility to yield giant optical anisotropy. Finally, the pronounced optical anisotropy in $CuAlO_2$ highlights its potential applications as an optical component in a wide spectrum. For example, its pronounced $|\Delta n(\omega)|$ is ideal for devices such as electro-optic modulators and nonlinear optics[53] in UV-visible regions. These unique properties position $CuAlO_2$ as a promising material for emerging optoelectronic and photonic applications.

**Acknowledgment**

This research was supported by National R&D Program through the National Research Foundation of Korea (NRF) funded by Ministry of Science and ICT (2022M3H4A1A04096465, RS-2024-00348920, 2023R1A2C1007634, 2021M3H4A1A0305486411, RS-2023-00257666, RS-2023-00210295, NRF-2021R1A5A1032996, and NRF-2021R1C1C1010216). This research was supported by Korea Basic Science Institute (National research Facilities and Equipment Center) grant funded by the Ministry of Education (No. RS-2024-00435344). M-2000 ellipsometer (J.A.Woolam Co.) for optical measurements was supported by the IBS Center for Correlated Electron Systems, Seoul National University.

**Contributions**

B.K., J.M.O., H.J., and C.S. conceptualized this work. M.K. and J.M.O. synthesized and characterized the single crystals. B.K. and C.S. performed spectroscopic ellipsometry. J.S. and H.J. conducted the first-principles calculations. B.K., M.K., U.C., U.S., K.C., J.M.O., and C.S. analyzed the experimental data. B.K. and C.S. wrote the paper with input from all coauthors.

**Ethics declarations**

Competing interests

The authors declare no competing interests.

**Data availability**

All data to evaluate the conclusions are present in the manuscript, and the Supplementary Material. Raw data are available from the corresponding authors on request.

**Code availability**

The codes of this study are available from the corresponding authors upon reasonable request.

**Methods**

Sample preparation and characterization

A high-quality single crystal of $CuAlO_2$ was grown by the reactive crucible melting method[18]. A powder of $Cu_2O$ (2.5 g) was placed into an alumina crucible (99.9% purity), which was located inside a box furnace and heated to a temperature of 1225°C and held for 6 hours. Following the holding process, the furnace was cooled to 1150°C at a rate of 1°C per hour and then cooled to room temperature. A number of crystals were formed within the crucible and could be separated mechanically to obtain a single crystal. The obtained crystals were 1–2 mm in length and of a black, shiny, hexagonal pillar shape. The crystal structure was confirmed by single-crystal X-ray diffraction. The measurements were performed on a D8 Venture (Bruker) using a Mo K$\alpha$ X-ray source. The crystal structure was determined by refinement using the built-in software (APEX 4).

Ellipsometry and optical conductivity of $CuAlO_2$

The dielectric function of $CuAlO_2$ was obtained using an M-2000 ellipsometer (J. A. Woolam Co.). All samples were mounted with silver epoxy onto oxygen-free copper cones to eliminate reflections from the sample's backside. The ellipsometry parameters, $\Psi$ and $\Delta$, of $CuAlO_2$ were measured over the energy range of 0.74 to 6.46 eV (5900 to 52000 cm$^{-1}$) at 70° incident angles $\theta$. Due to the large size and sufficient thickness of the $CuAlO_2$ single crystals, we were able to independently measure the sample's top surface (normal vector is parallel to the $c$-axis) and side shiny surface (normal vector is perpendicular to the $c$-axis). Despite the anisotropic nature of the material, the optical axis ($c$-axis) is either parallel or perpendicular to the $s$-polarization in both the top and side views, resulting in no off-diagonal dielectric tensor components. This means that the contribution is largely from the direction parallel with $s$-polarization[54]. While the refractive index for other axes is included, the insensitivity to this axis in reflectivity geometry allows us to effectively ignore these contributions[5]. Therefore, the ellipsometry parameters obtained from each surface are dominantly governed by the dielectric function along the parallel direction to the $s$-polarization. The dielectric functions, $\varepsilon_1$ and $\varepsilon_2$, were simultaneously extracted through transformation process, $\varepsilon_1 + i\varepsilon_2 = \sin^2\theta[1+\tan^2\theta(1-\rho/1+\rho)^2]$, where $\rho = \tan(\Psi)e^{i\Delta}$. Various optical constants are calculated from dielectric function. For low-temperature measurements, we calibrated the window effect and determined the $\Delta$ offset using a 25 nm $SiO_2$/Si wafer. To prevent ice formation on the sample surface, the chamber was baked to achieve a base pressure below $5 \times 10^{-9}$ Torr.

Electronic structure calculations

Density functional theory calculations were performed with the projector-augmented wave method as implemented in the Vienna Ab-initio Simulation Package (VASP) code[55,56]. The potential was generated with the valence electron configurations: Cu $3p^6$ $3d^{10}$ $4s^1$, Al $3s^2$ $3p^1$, and O $2s^2$ $2p^4$. The exchange-correlation function was described as the GGA of the Perdew-Burke-Ernzerhof (PBE)[57]. Plane-wave cutoff energy was 700 eV, and the first Brillouin zone was sampled with an 11×11×11 Monkhorst–Pack k-point grid[58]. The energy convergence for the self-consistent field approach was set to $10^{-6}$ eV. Structural optimization continued until the forces on each atom were reduced to less than 0.002 eV/A.

[†]These authors contributed equally.
[*]chsohn@unist.ac.kr

*hsjin@unist.ac.kr

*okjongmok@pusan.ac.kr

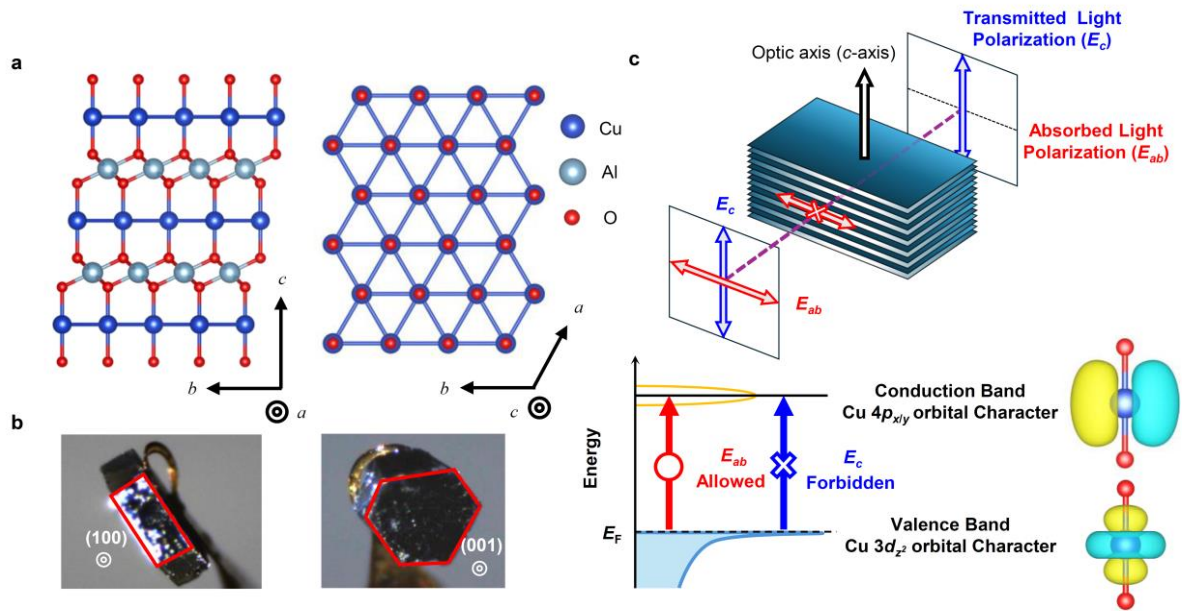

**Figure 1. Schematics of three-dimensional crystal structure and optical anisotropy of CuAlO$_2$**

**a**, Schematic representation of the lattice structure of CuAlO$_2$. Blue, light blue, and red spheres indicate Cu, Al, and O atoms, respectively. **b**, Microscopy images of a CuAlO$_2$ single crystal. The left panel shows the side view with the normal vector along (100) direction, while the right panel displays the top view with the normal vector along (001). The red box illustrates the plane perpendicular to each normal vector. The black color of the crystal is attributed to 1.4 % of Cu$^{2+}$ impurity due to the sufficient thickness[18]. **c, (Upper)** Schematic illustrating of the optical anisotropy of CuAlO$_2$. The red and blue arrows indicate the polarization of incident light, while the purple dashed line represents the direction of incident light propagation. The optic axis (*c*-axis) is shown by the black arrow. Light polarized parallel to the *c*-axis ($E_c$) is transmitted through the *ac*-plane (Left panel of **Fig. 1b**), whereas light polarized parallel to the *ab*-plane ($E_{ab}$) is absorbed. **(Lower)** Schematic of the band structure and corresponding orbital character of CuAlO$_2$. The highest valence band is predominantly characterized by Cu $3d_{z^2}$, while the lowest conduction band features Cu $4p_{x/y}$ orbitals. The colossal optical anisotropy in CuAlO$_2$ is explained by the selection rules governing the transition between these orbitals and the polarization of incident light.

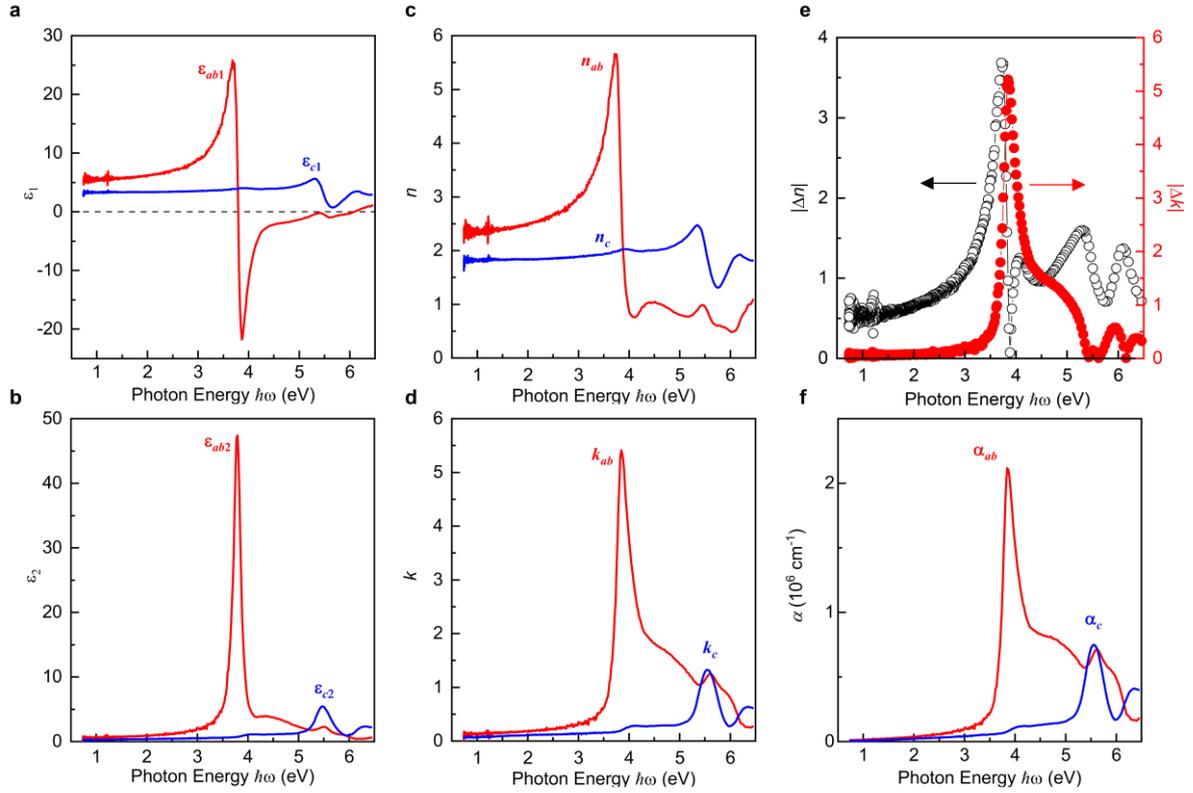

**Figure 2. Giant optical anisotropy of CuAlO$_2$**
**a,** Real part ($\varepsilon_1(\omega)$) and **b,** imaginary part ($\varepsilon_2(\omega)$) of the dielectric function of CuAlO$_2$ at room temperature. The red and blue lines represent the component for the *ab*-plane ($\varepsilon_{ab1}(\omega)$, $\varepsilon_{ab2}(\omega)$) and *c*-axis ($\varepsilon_{c1}(\omega)$, $\varepsilon_{c2}(\omega)$) components, respectively. The horizontal dashed line indicates $\varepsilon_1 = 0$. The dielectric functions were directly obtained from the ellipsometry parameters without model fitting. **c,** Refractive index ($n(\omega)$) and **d,** extinction coefficient ($k(\omega)$) of CuAlO$_2$ at room temperature. The red and blue lines represent the *ab*-plane ($n_{ab}(\omega)$, $k_{ab}(\omega)$) and *c*-axis ($n_c$ $k_c$) components, respectively. **e,** Absolute values of birefringence $|\Delta n(\omega)|$ and linear dichroism $|\Delta k(\omega)|$ of CuAlO$_2$ at room temperature. The black solid circles and red open squares denote $|\Delta n(\omega)|$ and $|\Delta k(\omega)|$, respectively. The maximum values of birefringence ($|\Delta n_{\max}|$) and linear dichroism ($|\Delta k_{\max}|$) are about 3.67 at 333 nm and 5.21 at 322 nm, respectively—representing the highest values reported to date. **f,** Optical absorption coefficient $\alpha(\omega)$ of CuAlO$_2$ at room temperature. The red and blue lines correspond to the in-plane ($\alpha_{ab}(\omega)$) and out-of-plane ($\alpha_c(\omega)$) components, respectively.

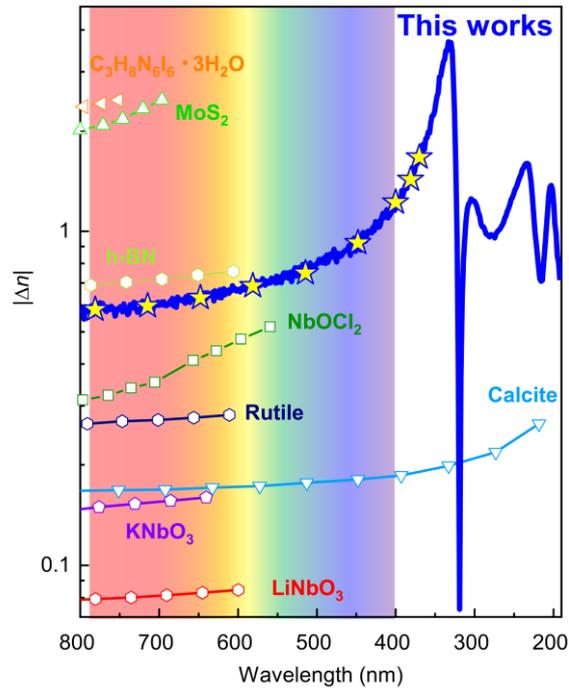

**Figure 3. Comparison of birefringence with other anisotropic materials**
The comparison of $|\Delta n(\omega)|$ between $CuAlO_2$ and other well-known anisotropic materials possessing transparency in the visible range[5,7,20-25]. The symbols represent the transparent regions, where the absorption coefficient deviates from the Urbach tail below the bandgap for each material[26]. The rainbow-shaded region highlights the visible range (400–780 nm). Notably, $CuAlO_2$ demonstrates both an exceptionally wide transparent range and the highest birefringence across the visible and ultraviolet regions, outperforming commercially available birefringent materials such as rutile, calcite, and $LiNbO_3$.

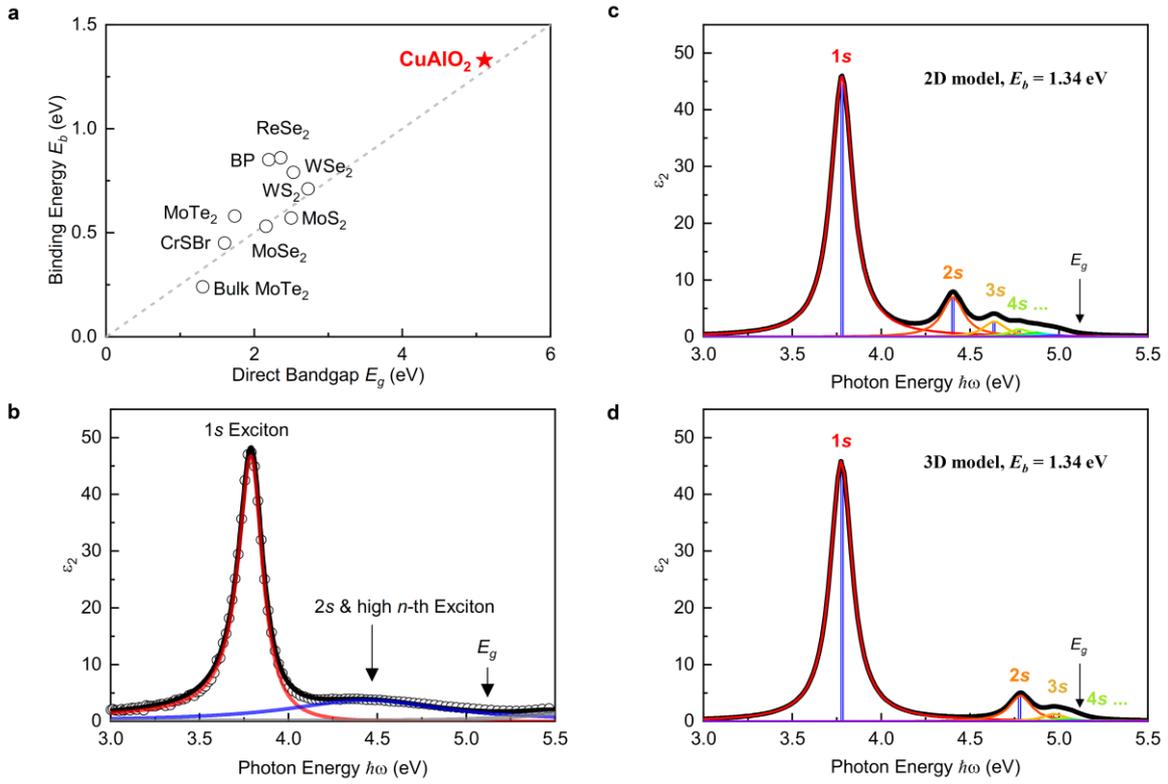

**Figure 4. Two-dimensionally confined exciton as the origin of colossal optical anisotropy**
**a**, Experimental observation of the electronic bandgap ($E_g$) and binding energy ($E_b$) in various two-dimensional materials and $CuAlO_2$. The red star and black circle represent the results for $CuAlO_2$ from this work and other materials[28-38], respectively. The gray dashed line marks the $E_b/E_g = 1/4$ scaling behavior, characteristic of the two-dimensional exciton. **b**, Experimental data and Fano-Lorentz fitting of $\varepsilon_{ab2}(\omega)$. The open circles indicate experimental data while colored solid lines exhibit the Fano-Lorentz fitting. The broad continuum observed above the 1s exciton is attributed to contributions from 2s and higher order n-th exciton states. The $E_g$ is obtained from Penn gap fitting. The calculated n-th series of excitons and simulated $\varepsilon_{ab2}(\omega)$ with $E_b = 1.34$ eV using the **c,** two-dimensional screened hydrogen model, and **d,** three-dimensional isotropic hydrogen model. The blue bars represent the calculated peak positions and relative oscillator strengths of the n-th exciton series for both models. The black line shows cumulative $\varepsilon_{ab2}(\omega)$, while the colored lines show the contribution from individual n-th exciton for both models. Here, we assume that all n-th exciton has identical linewidth with 1s exciton. The 2D model successfully captures the broad continuum observed at 4.47 eV in **b**, aligning well with the experimental results.

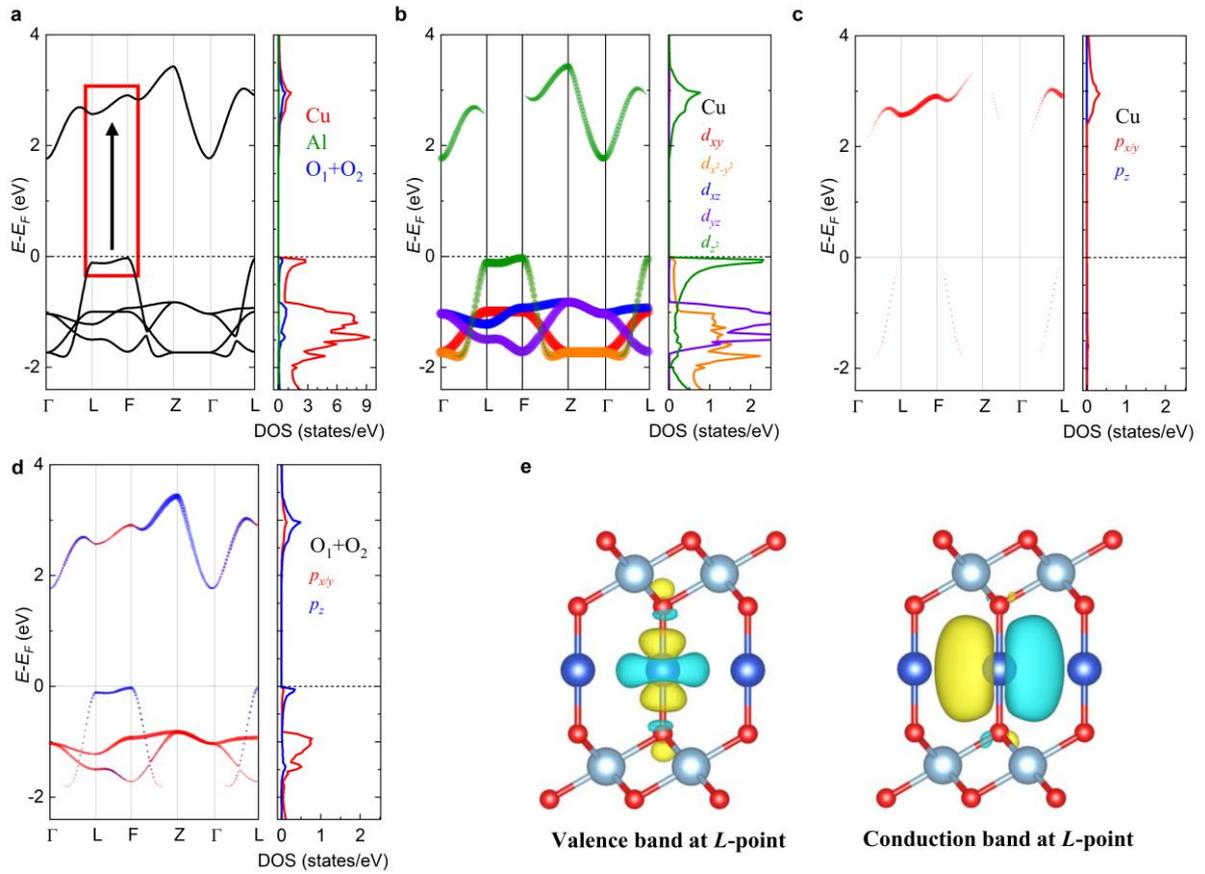

**Figure 5. Theoretical investigation of the orbital characters in CuAlO$_2$**
**a,** Band structure and atom-selective partial density of states (PDOS) of CuAlO$_2$ near the Fermi level. The results are qualitatively consistent with previous research[12,40]. Since the Al atom has negligible contributions near the Fermi level, the analysis focuses on Cu and O atoms. The orbital-projected band structures and PDOS of CuAlO$_2$ are represented by **b,** Cu $d$-orbitals, **c,** Cu $p$-orbitals, and **d,** oxygen $p$-orbitals. The color and size of the symbols indicate the orbital symmetry and relative intensity of the contributions, respectively. The band structure is plotted along the high-symmetry path in the rhombohedral Brillouin zone. **e,** Real-space projected Bloch functions for the highest valence band and lowest conduction band at $L$-point. Both functions are predominantly localized within the Cu layers, emphasizing the quasi-two-dimensional nature of the electronic states.

Supplementary Information for

# Colossal optical anisotropy in wide-bandgap semiconductor CuAlO$_2$

Baekjune Kang,[1,†] Junhee Shin,[1,†] Myeongjun Kang,[2] Uksam Choi,[1] Uihyeon Seo,[1] Kunook Chung,[1] Jong Mok Ok,[2,*] Hosub Jin,[1,*] and Changhee Sohn[1,*]

[1]Department of Physics, Ulsan National Institute of Science and Technology, Ulsan, 44919, Republic of Korea

[2]Department of Physics, Pusan National University, Pusan, 46241, Republic of Korea

# Contents

**Supplementary Note 1. Three-dimensional crystal and structural information of CuAlO$_2$**

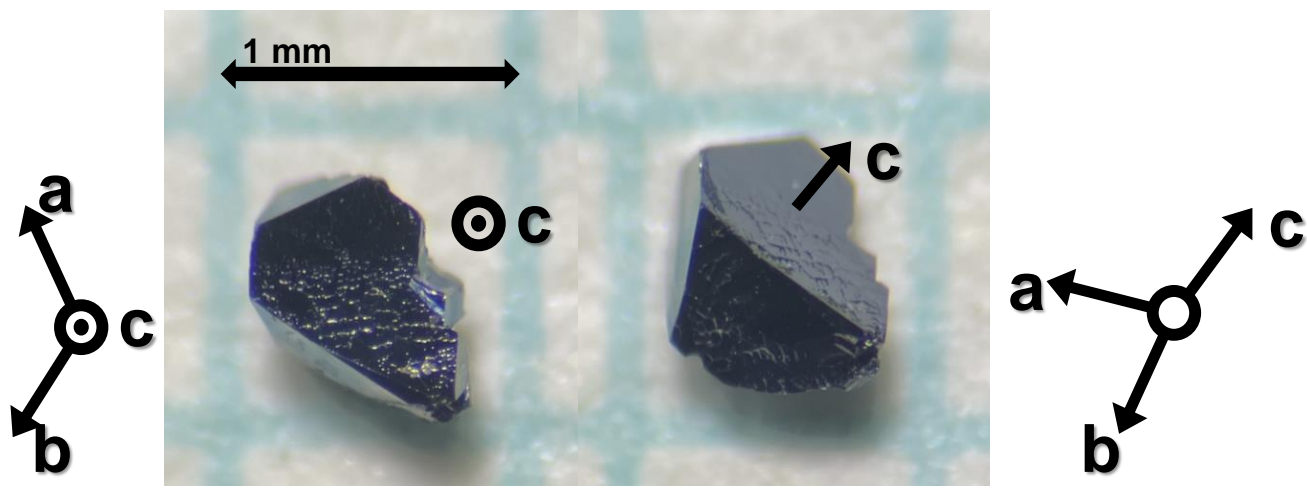

**Figure S1.** Microscopy images of CuAlO$_2$ single crystals along different crystallographic axes. The mm-sized crystals with sufficient thickness allow for precise measurement of optical anisotropy.

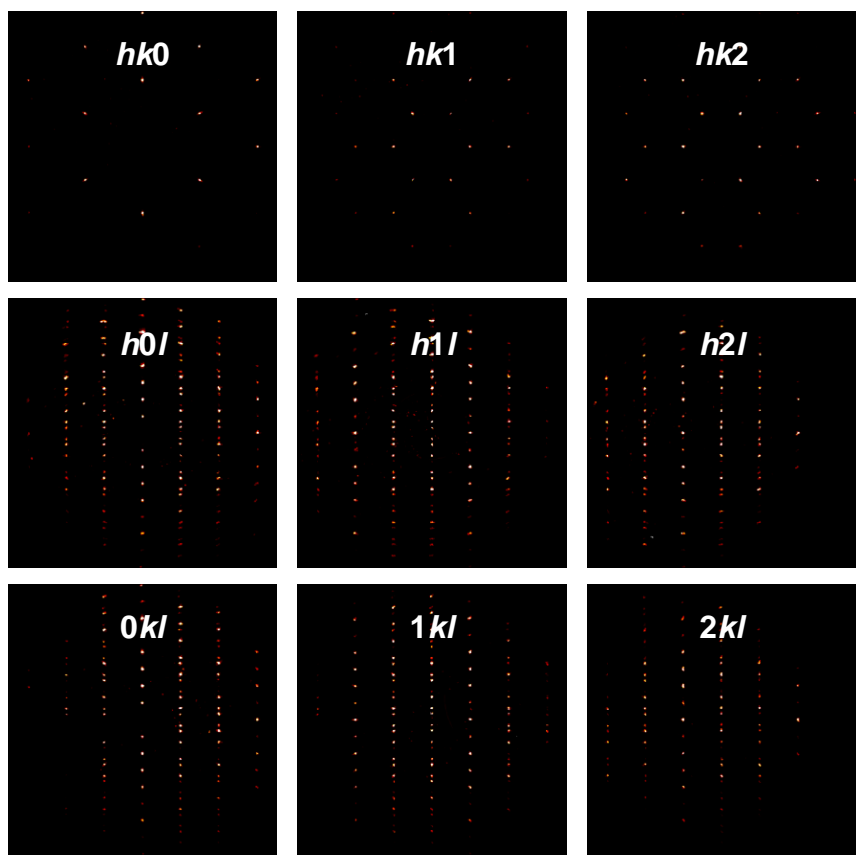

**Figure S2.** Single crystal X-ray diffraction patterns of CuAlO$_2$ measured along various crystallographic directions. The images correspond to reciprocal lattice planes labeled as *hk*0, *hk*1, *hk*2 (top row), *h*0*l*, *h*1*l*, *h*2*l* (middle row), and 0*kl*, 1*kl*, 2*kl* (bottom row).

In this note, we show a three-dimensional crystal and conduct a quantitative evaluation of the lattice structure of $CuAlO_2$. Due to the sufficient thickness of crystal shown in Fig. S1, we can conduct the single crystal X-ray diffraction of various direction of crystals. Figure S2 shows the single crystal X-ray diffraction of $CuAlO_2$. The quantitative information of lattice structure from the Rietveld refinement is listed in Table S1.

| Chemical formula | $CuAlO_2$ |
|---|---|
| Crystal system | Trigonal |
| Space group | $R\bar{3}m$ |
| Lattice parameter $a$ | 2.8537(6) Å |
| Lattice parameter $c$ | 16.926(4) Å |
| Volume | 119.37(7) Å$^3$ |
| Z | 3 |

**Table. S1.** Lattice structure of $CuAlO_2$ from the single crystal X-ray diffraction

**Supplementary Note 2. Experimental geometry and ellipsometry parameter**

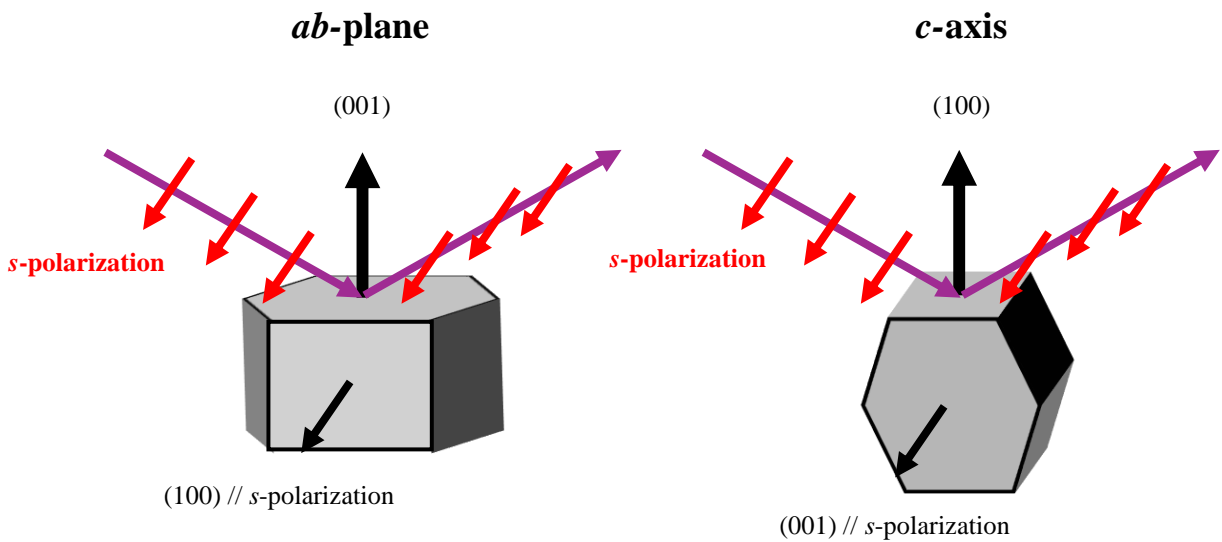

**Figure S3**. Experimental geometry for obtaining the ellipsometry parameters of $CuAlO_2$ along the (**Left**) *ab*-plane and (**right**) *c*-axis. The black arrows indicate the crystallographic lattice vectors, while the purple and red arrows represent the incident light and the *s*-polarization orientation, respectively. In this geometry, the optical constants parallel to the direction of *s*-polarization are dominant, while contributions from other directions are negligible.

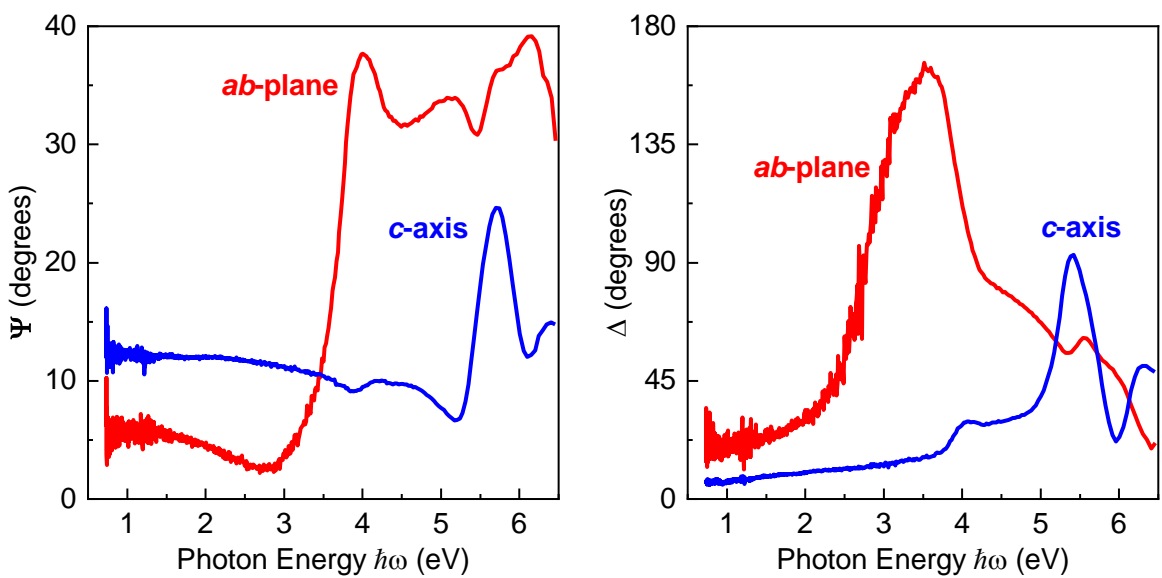

**Figure S4.** Experimentally obtained ellipsometry parameters, (**Left**) $\Psi$ and (**Right**) $\Delta$, for $CuAlO_2$ at 300 K, ambient pressure with an incident angle of 70 degrees along different crystallographic orientation. Red and Blue lines indicate the *ab*-plane and *c*-axis, respectively.

To investigate the anisotropic optical properties of $CuAlO_2$, we conducted ellipsometry experiments to measure its complex optical constants. Ellipsometry determines Ψ (the amplitude ratio of reflected *p*- and *s*-polarized light) and Δ (the phase difference between them) to extract optical constants. The single crystal nature and sufficient thickness of $CuAlO_2$ allows for the direct determination of these constants without relying on model fitting, a significant advantage over previous studies.

For anisotropic materials, reflected light often includes cross-polarized components, complicating analysis. However, aligning the optical axis ((001)-direction) parallel or perpendicular to the *s*-polarization eliminates these cross components, enabling independent measurement of the parallel axis to *s*-polarization[54]. This setup provides an information about the axis parallel to *s*-polarization.

Figure S3 illustrates the experimental geometry, where the *s*-polarized light is aligned parallel to specific crystallographic axes. This configuration enables independent measurements of optical parameters for *ab*-plane and *c*-axis directions, corresponding to cases where the (100) and (001) axes are parallel with the *s*-polarized light while (001) and (100) is parallel (perpendicular) to axis of reflection (*s*-polarization). For simplicity, only the *s*-polarized light configuration is shown. Using this geometry, we obtained ellipsometry parameters for each crystallographic axis (shown in Fig. S4). These parameters were converted directly into dielectric functions, as described in the Methods section. Unlike previous studies that relied on fitting to anisotropic models, our direct experimental approach provides independent and accurate measurements of $CuAlO_2$'s anisotropic properties.

# Supplementary Note 3. Absorption and transparent region of $CuAlO_2$

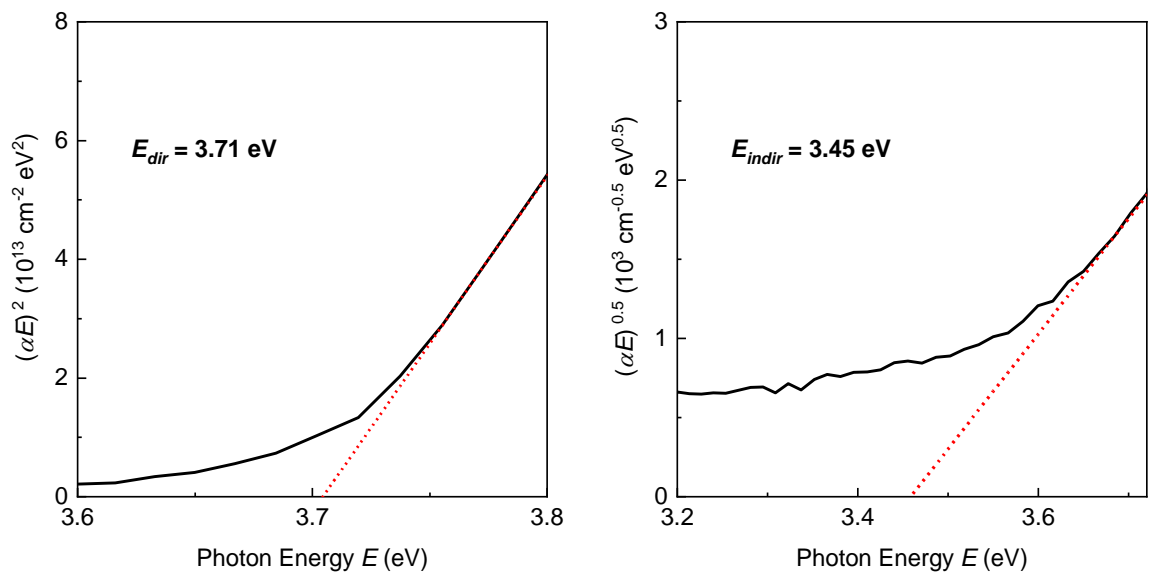

**Figure S5.** $(\alpha E)^n$ versus $E$ graphs for $CuAlO_2$ used to determine the direct and indirect bandgaps, where $\alpha$ and $E$ are absorption coefficient and photon energy, respectively. The black solid lines and red dashed lines indicates the $(\alpha E)^n$ and extrapolation of linear function at maximum slope points, respectively. **(Left)**, Plot for the direct bandgap $E_{dir}$ with $n = 2$, yielding an $E_{dir}$ of 3.71 eV. **(Right)**, Plot for the indirect bandgap $E_{indir}$ with $n = 0.5$, yielding an $E_{indir}$ of 3.45 eV.

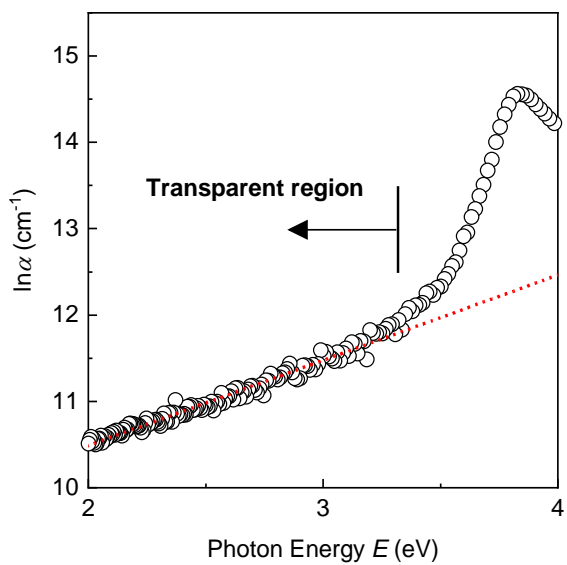

**Figure S6.** Plot of $\ln \alpha$ versus $E$ for $CuAlO_2$. The black open circles represents the experimental data, while the red dashed line indicates the Urbach fitting below bandgap.

In this note, we estimate the direct bandgap ($E_{dir}$), indirect bandgap ($E_{indir}$), and Urbach tail of $CuAlO_2$. Previous studies have shown that $CuAlO_2$ has a $E_{dir}$ of 3.5 eV and an $E_{indir}$ of 3.0 eV. To determine these values, we constructed plots of $(\alpha E)^n$ versus $E$, where $\alpha$ is the absorption coefficient, $E$ is the photon energy, and $n$ is an exponent constant. We set $n = 2$ for the $E_{dir}$ and $n = 0.5$ for the $E_{indir}$. Figure S5 presents the $(\alpha E)^n$ versus $E$ plots, with extrapolated linear functions derived from the points of maximum slope. Using this approach, we identified $E_{dir} = 3.71$ eV and $E_{indir} = 3.45$ eV. However, a slight amount of absorption is observed even below these bandgap energies.

This additional absorption can be attributed to the Urbach tail which represents defect-induced absorption below the optical bandgap typically following an exponential dependence on photon energy[26]. Figure S6 shows the $\ln \alpha$ as a function of $E$, revealing a linear behavior from 1 eV to 3.35 eV, indicating that disorder dominates absorption in this range. A $Cu^{2+}$ defect concentration of around 1.4% could account for this absorption[18]. Therefore, we conclude that above 3.35 eV, absorption is intrinsic to $CuAlO_2$, while below 3.35 eV, it primarily arises from defect-induced absorption. In other words, $CuAlO_2$'s inherent transparency is allocated until this threshold.

**Supplementary Note 4. The Penn gap fitting of $CuAlO_2$**

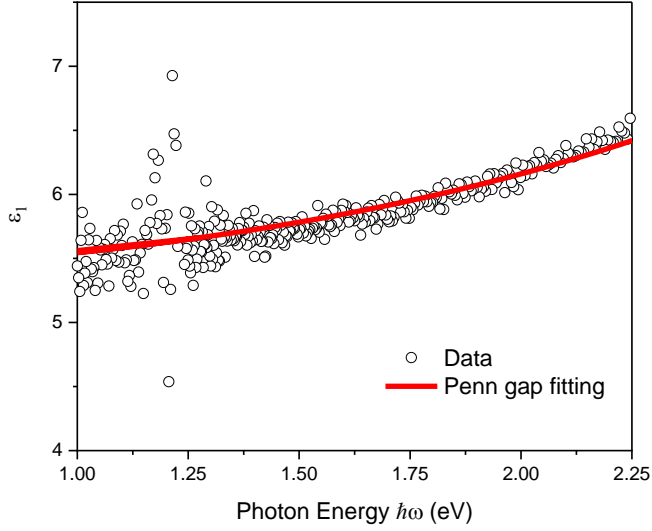

**Figure S7.** Experimental data and Penn gap fitting of $\varepsilon_{ab1}(\omega)$ at 300 K. The black open circles and red solid line indicate the experimental data, Penn gap fitting, respectively.

In this note, we conduct a quantitative evaluation of the dielectric function to investigate the electronic bandgap in $CuAlO_2$. Previous studies estimated that the intense optical transition (Penn gap) in $CuAlO_2$ occurs around 5 eV[16,17]. However, our results indicate a strong transition at a significantly lower energy of 3.78 eV. This indicates that additional excitonic contributions, which have been recently observed in experiments, play a crucial role in the observed transition. To describe the behavior of the dielectric function in the visible range sufficiently lower than bandgap, which lies below the 1s exciton transition (< 2.25 eV), we use the following equation for the Penn gap[39]:

$$\varepsilon_1(\omega) = 1 + (\varepsilon_\infty - 1)\frac{\omega_P^2}{\omega_P^2 - \omega^2} \qquad \text{Eq. (1)}$$

Here, $\varepsilon_\infty$ and $\omega_P$ represents the background dielectric function beyond the phonon region and Penn gap, respectively. Figure S7 shows experimental data and fitting with Penn gap model. The fitting result is $\varepsilon_\infty = 5.37$ and $\omega_P = 5.12$ eV, which is consistent with previous experimental fitting and theoretically calculated bandgap from sc-GW methods[40], the most precise calculation for estimate the bandgap of wide bandgap semiconductor.

**Supplementary Note 5. Fano-Lorentz fitting of CuAlO$_2$**

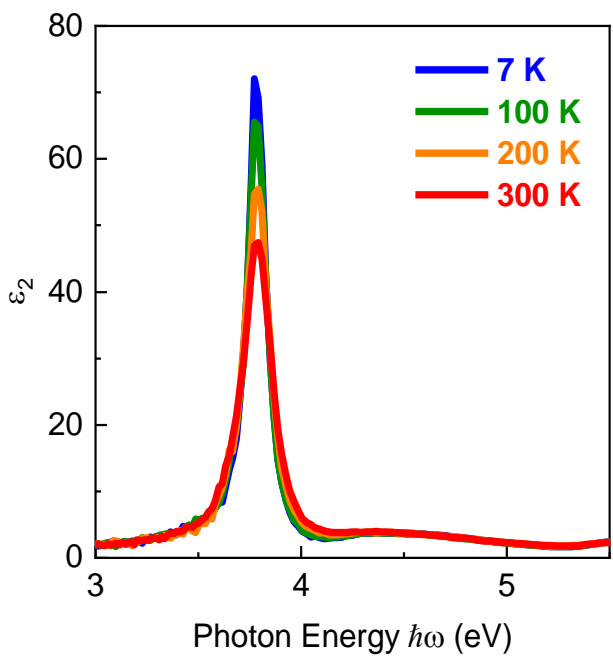

**Figure S8.** Temperature dependent $\varepsilon_{ab2}(\omega)$ as a function of photon energy. By decreasing temperature, the Fano-like behavior at 4.15 eV is pronounced while peak becomes sharpen, reaching 120 meV linewidth.

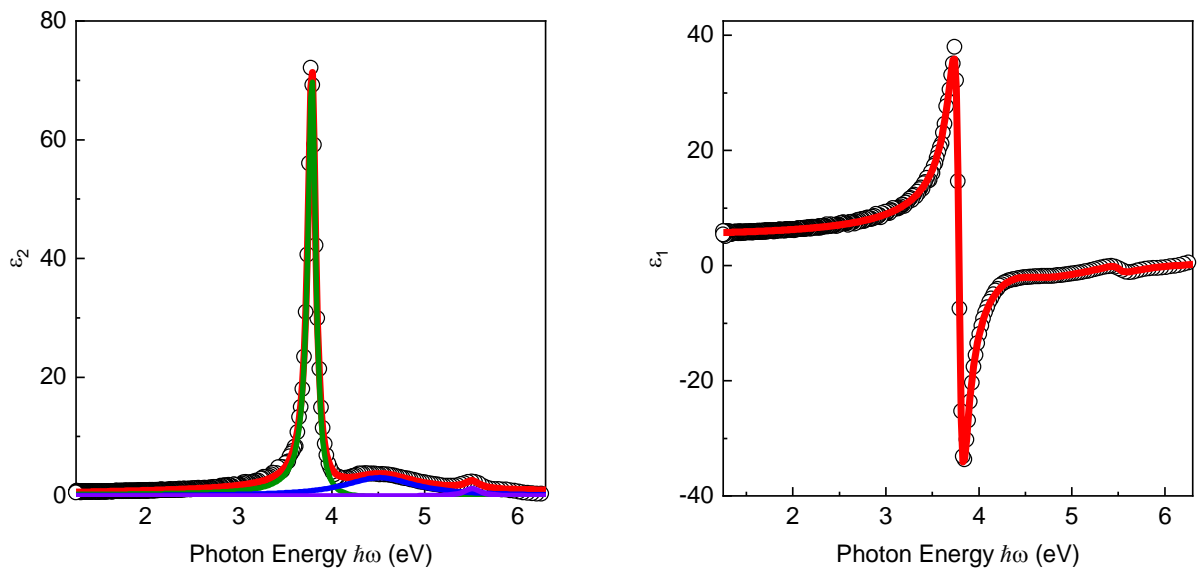

**Figure S9.** Experimental data and simultaneous fitting of **(Left)** $\varepsilon_{ab2}(\omega)$ and **(Right)** $\varepsilon_{ab1}(\omega)$ at 7 K. The black open circles represent the experimental data, while the red solid line shows the sum of all contributions for both $\varepsilon_{ab1}(\omega)$ and $\varepsilon_{ab2}(\omega)$. The green solid line represents the Fano function, while the blue and navy solid lines correspond to Lorentz functions in $\varepsilon_{ab2}(\omega)$.

In this note, we apply Fano-Lorentz fitting to explore the temperature-dependent behavior of exciton in $CuAlO_2$, particularly due to the asymmetric nature of the exciton peak. While most interband transitions are well described by a Lorentz function, the exciton peak in the in-plane dielectric function $\varepsilon_{ab2}(\omega)$ of $CuAlO_2$ shows noticeable asymmetry, with a sharper decline on the high-energy side. At low temperatures, this peak takes on a Fano-like profile, as shown in Fig. S8. The coupling between 1$s$ exciton and higher-order exciton would be responsible to the Fano resonance. To capture the asymmetric exciton peak and its temperature dependence in $CuAlO_2$, we applied a Fano-Lorentz fitting model. Figure S9 shows the fitting of $\varepsilon_{ab1}(\omega)$ and $\varepsilon_{ab2}(\omega)$ at 7 K using one Fano function and two Lorentz functions. The Fano term is assigned to 1$s$ exciton while broad Lorentz term at 4.47 eV is contributed from 2$s$/higher-order exciton series.

**Supplementary Note 6. Effective medium approximation for CuAlO$_2$**

In this note, we utilize the effective medium approximation to obtain quantitative insights into the two-dimensional excitons in CuAlO$_2$. This approach is chosen for its ability to provide simplified yet insightful information about two-dimensional excitons. While quantitative details about excitons can be obtained using Bethe-Salpeter Equation calculations with a nonlocal potential, such method is computationally intensive and beyond the scope of our study. Instead, we adopt the framework from Ref. [45], which offers an analogous method for extracting two-dimensional exciton information similar to the established framework for three-dimensional excitons[45].

We assume CuAlO$_2$ as a superlattice consisting of Cu$^+$ and AlO$_2^-$ layers, each with distinct dielectric constants. For simplicity, we assume the dielectric constants of each layer are isotropic, with $\varepsilon_{Cu}$ representing the Cu$^+$ layers and $\varepsilon_{Al}$ representing the AlO$_2^-$ layers. The thickness of the Cu$^+$ layer is taken as half the length of the O-Cu-O dumbbell (2.038 Å), while the remaining thickness corresponds to the AlO$_2^-$ layer (3.622 Å). The normalized thicknesses, relative to the total superlattice thickness, are $d_1 = 0.36$ for the Cu$^+$ layer and $d_2 = 0.64$ for the AlO$_2^-$ layer. We then apply the effective medium approximation to calculate the overall dielectric constant of the superlattice, which can be expressed as:

$$\varepsilon_{\parallel} = d_1\varepsilon_{Cu} + d_2\varepsilon_{Al}$$

$$\frac{1}{\varepsilon_{\perp}} = \frac{d_1}{\varepsilon_{Cu}} + \frac{d_2}{\varepsilon_{Al}}$$

Eq. (2)

To apply the model, we define the appropriate dielectric constant based on the experimental value at the lowest measured energy (0.74 eV). Solving the equation yields two possible values for the dielectric constant of the Cu$^+$ layer, depending on whether $\varepsilon_{Cu}$ is larger or smaller than $\varepsilon_{Al}$.

To select the appropriate solution for the dielectric constant, we compare the AlO$_2^-$ layer in CuAlO$_2$ to Al$_2$O$_3$, as both materials share similar structural features. Both materials have octahedral structures composed of aluminum and oxygen atoms, with the *c*-axis corresponding to the <111> direction of the octahedron. And the length of Al-O of CuAlO$_2$ and Al$_2$O$_3$ is 1.832 Å and 1.855 Å, respectively. The structural similarity provides the similar dielectric properties. This comparison allows us to estimate the dielectric function at lower energies. At 0.74 eV—well below the interband transition energy and above the phonon energy—, we expect the dielectric constants of both materials to be similar. The dielectric function of Al$_2$O$_3$ in this range is known to be approximately 2.7. Thus, it is reasonable to conclude that the dielectric constant of the AlO$_2^-$ layer is smaller than that of the Cu$^+$ layer, based on their comparable structures and expected behavior. In fact, when the dielectric constant of the Cu$^+$ layer is assumed to be larger, we obtain $\varepsilon_{Cu} = 11.0154$ and $\varepsilon_{Al} = 2.3351$—values that resemble the previously reported dielectric constant of Al$_2$O$_3$.

## Supplementary Note 7. The quantitative information of higher-order *n*-th exciton

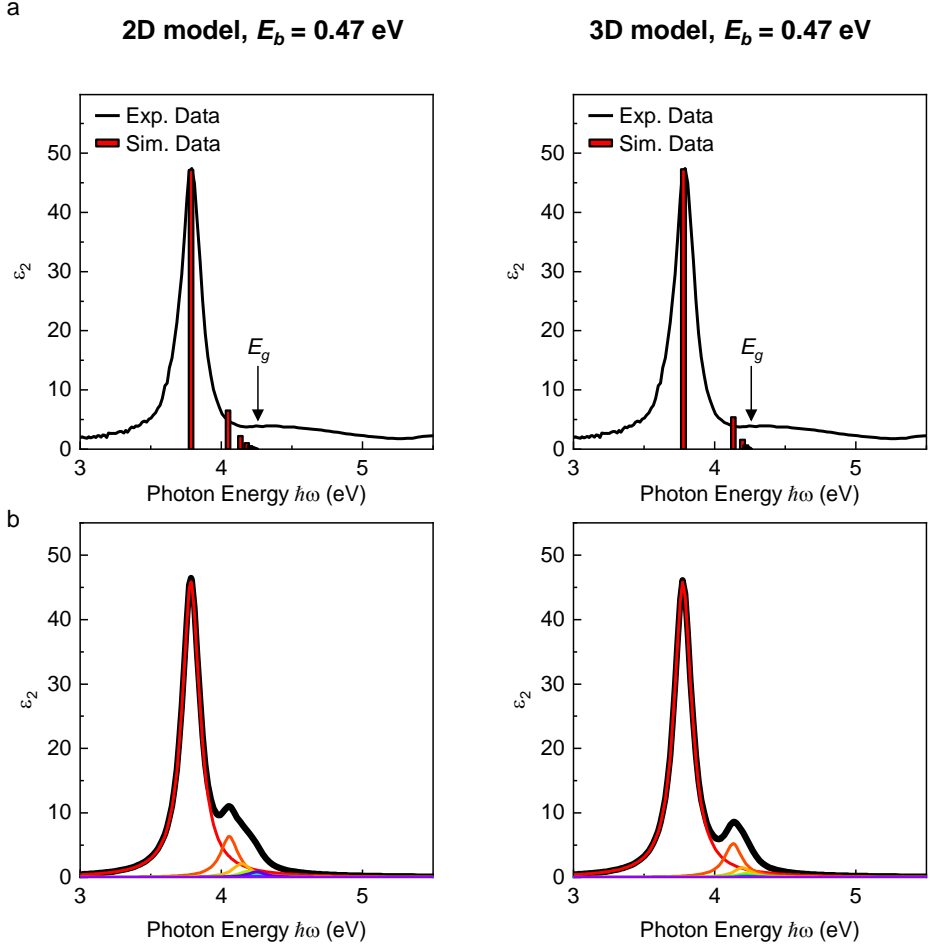

**Figure S10 a,** Comparison of experimental $\varepsilon_{ab2}(\omega)$ spectra and higher-order *n*-th exciton series for CuAlO$_2$ using the **(Left)** two-dimensional screened hydrogen model and **(Right)** three-dimensional hydrogen model with binding energy 0.47 eV. The black solid lines indicates the experimental data, while red bars provide peak position and relative oscillator strength of n-th exciton series for both models. **b,** Simulated $\varepsilon_{ab2}(\omega)$ with the same broadening applied to each *n*-th exciton for **(Left)** two-dimensional screened hydrogen model and **(Right)** three-dimensional isotropic hydrogen model. The black line represents cumulative $\varepsilon_{ab2}(\omega)$, while the colored lines show the contribution from *n*-th exciton.

In this note, we quantitatively investigate the higher-order *n*-th exciton series in CuAlO$_2$. For an ideal two-, and three- dimensional system, the *n*-th exciton can be described using the two- (three-) dimensional hydrogen model, with the binding energy $E^{2D}_n$ ($E^{3D}_n$) and oscillator strength $f^{2D}_n$ ($f^{3D}_n$) defined as follows[41,46]:

$$E_n^{2D} = -\frac{\mu}{2\left(n-\frac{1}{2}\right)^2}, \qquad E_n^{3D} = -\frac{\mu}{2n^2}$$

$$f_n^{2D} = -\frac{A}{\omega_n \left(n-\frac{1}{2}\right)^3}, \qquad f_n^{3D} = -\frac{A}{\omega_n n^3}$$

Eq. (3)

Here, $\mu$, $A$, and $\omega_n$ represent the effective mass of exciton, amplitude constant, and peak position of $n$-th exciton, respectively. The three-dimensional material case is effectively applied in real material. However, two-dimensional materials are embedded in a three-dimensional environment in real experimental geometry, which means the materials are surrounded by a dielectric background[44]. In such cases, the potential energy can no longer be described simply using a Coulomb potential, and although an analytical solution has been proposed, it remains highly complex[41]. However, recent studies on two-dimensional materials have revealed that by incorporating an effective dielectric contribution, the resulting solutions closely resemble those from the analytical models[44]. Therefore, to gain insight into the higher-order exciton series in $CuAlO_2$, we employ a simplified model that includes this effective dielectric contribution.

The model begins with the two-dimensional Schrödinger equation, which includes a two-dimensional Coulomb potential. The effective dielectric function of $n$-th exciton, $\varepsilon_n$, experienced by the material is integrated into the equation, leading to the following expression:

$$-\frac{\hbar^2}{2\mu}\left(\frac{\partial^2}{\partial x^2}+\frac{\partial^2}{\partial y^2}\right)\varphi(x,y) - \frac{e^2}{4\pi\varepsilon_n(x^2+y^2)^{1/2}}\varphi(x,y) = E\varphi(x,y) \qquad \text{Eq. (4)}$$

, where $e$ and $\varphi$ are elementary charge and wavefunction of exciton. Solving this equation[45] gives the $E_n$ and $f_n$ for the exciton series:

$$E_n = -\frac{\mu}{2\varepsilon_n^2\left(n-\frac{1}{2}\right)^2}$$

$$f_n = -\frac{A}{\omega_n\varepsilon_n^2\left(n-\frac{1}{2}\right)^3} \qquad \text{Eq. (5)}$$

Since we know the $E_1$, we can derive the $E_n$ and $\omega_n$ for all $n$-th excitons, as long as the information on $\varepsilon_n$ is available. Furthermore, the relative intensity of each excitonic peak, compared to the first, can also be calculated. Recent studies have presented the $\varepsilon_n$ in two-dimensional materials when embedded in a three-dimensional dielectric environment. We obtained $\varepsilon_n$ via Ref. [44], $\varepsilon_{Cu}$, and $\varepsilon_{Al}$ for $n$-th order exciton. Using this approach, we have produced Fig. 4c, which illustrates the calculated results for all $n$-th excitons. However, to the best of our knowledge, no previous research has examined the quantitative scattering rate for these higher-order excitonic states.

Finally, we demonstrate that the binding energy obtained from the Penn gap is not an exaggerated value. Ref. [12] and Ref. [13] report different binding energies calculated via BSE calculation: 0.47 eV[12] and 1.2 eV[13], respectively. While the former deviates from our experimental results, the latter aligns closely with them. This deviation has been attributed to the pseudopotential used for the Cu atom in the band structure calculation. For comparison, we also calculated the exciton series for a binding energy of 0.47 eV. Figure S10 displays the peak positions and oscillator strengths of the $n$-th exciton series and the simulated $\varepsilon_{ab2}(\omega)$ for both models. Both models fail to reproduce the experimental spectrum, particularly because the high $n$-th exciton series overlaps with the Fano-like dip. Thus, a binding energy of 0.47 eV is insufficient to explain the experimental spectrum using either model.